\documentclass[twocolumn,showpacs,preprintnumbers,amsmath,amssymb]{revtex4}

\usepackage{graphicx}
\usepackage{graphics}

\usepackage{dcolumn}
\usepackage{bm}

\begin{document}

\preprint{preprint submitted to Journal of Applied Physics -\\
October 14, 2005}

\title{Magnetostrictive hysteresis of TbCo/CoFe multilayers and magnetic domains. }

\author{J.-Ph. \textsc{Jay}}
\email{jay@univ-brest.fr}
\author{F. \textsc{Petit}} 
\author{J. \textsc{Ben Youssef}}
\author{M.V. \textsc{Indenbom}}

\affiliation{%
Laboratoire de Magn\'etisme de Bretagne - CNRS FRE 2697\\
Universit\'e de Bretagne Occidentale -
\\6, Avenue le Gorgeu C.S.93837 - 29238 Brest Cedex 3 - FRANCE.\\
}

\author{A. \textsc{Thiaville}}
\author{J. \textsc{Miltat}}
\affiliation{
Laboratoire de Physique des Solides - CNRS UMR 8502\\
Universit\'e Paris Sud -
\\Batiment 510, 91405 Orsay Cedex - FRANCE.\\
 }

\date{Received}  

\begin{abstract}
Magnetic and magnetostrictive hysteresis loops of TbCo/CoFe
multilayers under field applied along the hard magnetization axis
are studied using vectorial magnetization measurements, optical
deflectometry and magneto optical Kerr microscopy. Even a very small
angle misalignment between hard axis and magnetic field direction is
shown to drastically change the shape of magnetization and
magnetostrictive torsion hysteresis loops. Two kinds of magnetic
domains are revealed during the magnetization: big regions with
opposite rotation of spontaneous magnetization vector and
spontaneous magnetic domains which appear in a narrow field interval
and provide an inversion of this rotation.

We show that the details of the hysteresis loops of our
exchange-coupled films can be described using the classical model of
homogeneous magnetization rotation of single uniaxial films and the
configuration of observed domains. The understanding of these
features is crucial for applications (for MEMS or microactuators)
which benefit from the greatly enhanced sensitivity near the point
of magnetic saturation at the transverse applied field.
\end{abstract}

\maketitle



\section{Introduction}

Spontaneous microscopic magnetostrictive deformations of a
magnetically ordered material can be transformed into its
macroscopic deformation by a modification of its magnetization
structure whose energy is much lower than the energy of an
equivalent elastic deformation.
Materials developing this property are thus very attractive for
actuator and sensor devices such as microrobots, micromotors
etc\cite{JAC-actuators},\cite{pasquale}. The main advantages of
magnetostrictive materials over piezoelectric or electrostrictive
materials is the capability of remote addressing
and controlling by an external magnetic field without direct electrical contacts. 

One of the key problems for practical applications of microsystems
is to reduce the magnetic driving field. The idea proposed by Quandt
\cite{Quandt-jap85},\cite{Quandt-jap81} was to combine the giant
magnetostrictive properties of  rare earth transition metal based
alloys (Terfenol like alloys : TbFe or TbCo) and the high
magnetization and soft properties of transition metal alloys (such
as CoFe) in multilayer films. In order to achieve this goal the
layers should be strongly coupled. Nevertheless, even in this case
it is not a priori clear whether the magnetic parameters will be a
simple average of those of each individual layer or a more complex
model should be used for the description of their properties.

Recent work \cite{tiercelin-jmmm249}  has shown that the well known
magnetic instability of uniaxial magnetic materials in the vicinity
of their saturation point under the magnetic field applied along the
hard magnetization axis  can be used to increase the sensitivity of
micro electromechanical systems (MEMS) based on TbFe/Fe multilayers.
This effect is sometimes presented as a spin reorientation phase
transition at the critical field equal to the anisotropy field as
was introduced in \cite{LandauL}. At the same time the observed
deformation of glass plates with TbFe/Fe multilayers deposited on
them in the desired magnetic field appeared to be much more complex
than expected. So, for further technical applications  a better
understanding of background mechanisms is necessary.

In this article we provide a detailed experimental study of
magneto-elastic and magnetization behavior of multilayers with giant
magnetostriction and combine them with observations of magnetic
domains {\it under the same conditions} using the magneto-optical
Kerr effect. The measurements are compared with a model taking into
account the coherent magnetization rotation under the applied field
of arbitrary direction and the role of the magnetic domains. We have
selected for the demonstration our results obtained on TbCo/CoFe
multilayers with very low saturation field $H_{sat}\approx 50$ Oe (4
 kA/m). Our experiments on TbFe/CoFe multilayers give practically
identical results.

\section{Experimental details}

Magnetic multilayers were grown onto  rectangular Corning glass
substrates (22 $\times$ 5 $\times$ 0.16 mm$^3$) from CoFe and TbCo
mosaic 4 inch targets using a Z550 Leybold RF sputtering equipment
with a rotary table technique. Base pressure prior to sputtering was
better than $4 \times 10^{-7}$ mbar. TbCo and CoFe were deposited
alternatively to get a  (TbCo/CoFe)$_{10}$ multilayer. TbCo was
deposited using 150 Watt RF power and argon gas pressure of $5
\times 10^{-3}$ mbar. The deposition conditions  for CoFe are
 200 Watt RF power and argon gas  pressure of $1
\times 10^{-2}$ mbar. Samples were deposited under a static field of
$\sim$ 300 Oe (24 kA/m) applied along the long side of the substrate
to favor an uniaxial magnetic anisotropy. No annealing treatment was
applied after deposition. The sample studied in this paper is:
\{[Tb$_{34\%}$Co$_{66\%}$]60 \AA/[Co$_{42\%}$Fe$_{58\%}$]50\AA\}
$\times$ 10. The chemical composition was determined on separately
prepared monolayer samples (TbCo and CoFe) using a X Ray
Fluorescence equipment. The deposition rates were calibrated using
Tencor profilometer on separate single layers. As usual for this
type of material  CoFe is polycrystalline while TbCo is amorphous as
proved by M\"{o}ssbauer spectroscopy \cite{Mossbauer-Neel2004}.

\begin{figure}[h]
\includegraphics[angle=0,width=8.5cm]{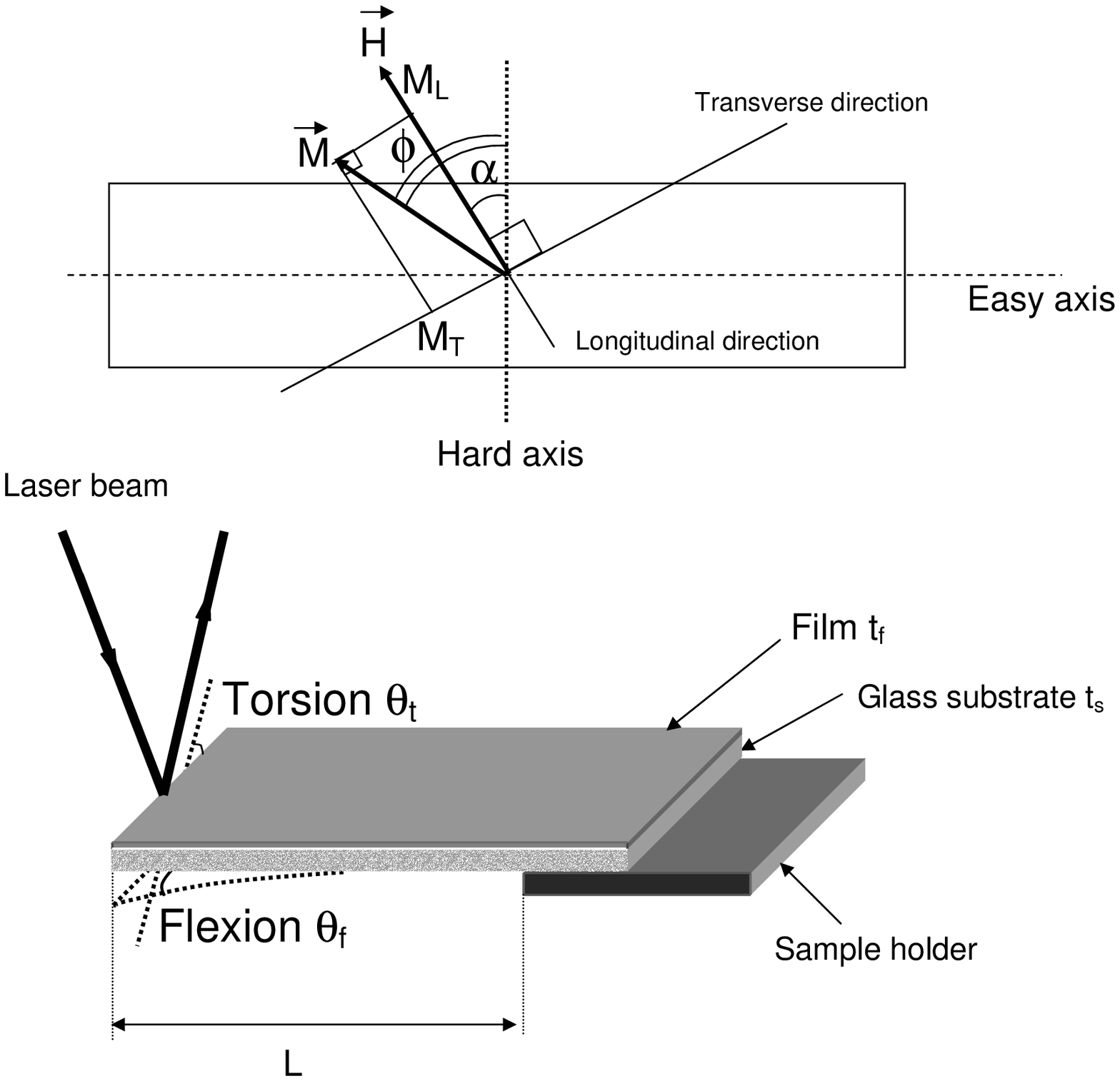}
\caption{\label{dispo}Geometry of the experiments: (upper panel)
Magnetic measurement; (lower panel) Measurements of magnetoelastic
deformations (flexion and torsion). Shown  downwards flexion is
defined as negative.}
\end{figure}

The  hysteresis loops were measured using a vibrating sample
magnetometer (VSM) that was modified in order to record the
evolution of both longitudinal and transverse components of the
sample magnetization. For this purpose two sets of detection coils
either parallel to the applied field (longitudinal component $M_L$)
or perpendicular to the applied field (transverse  component$M_T$)
were used. The vectorial measurements give direct information about
the direction of the magnetization rotation crucial for the
magnetoelastic behavior of the sample. The orientation $\alpha$ of
the magnetic field relatively to the hard magnetization axis of the
sample was varied by the rotation of the magnetometer head. Figure
\ref{dispo} presents a schematic diagram of the experiment.

 The magnetostrictive deformation of the samples was measured by
laser deflectometry,detecting  the flexion $\theta_f$ and torsion
$\theta_t$ angles at the free end of the plate.  The deflection of
the laser beam is detected using a two dimensional position
sensitive diode (PSD). The orientation of the principal axes of  PSD
relatively to the orientations of the laser beam displacement due to
torsion and flexion was carefully adjusted. The initial angles
$\theta_f$ and $\theta_t$, when there is no applied field and the
magnetization is oriented along the easy axis, are taken as zero.
The  orientation $\alpha$ of the applied field  was varied  by
rotation of the electromagnet independently of the optical
deflectometer.

The stresses in the magnetostrictive film produce curvature of the
sample with radius $R$ seen at its end as a deflection from the
initial horizontal plane, $\theta_f=L/R$, where $L$ is the "free"
sample length.
This flexion angle  $\theta_{f}$, measured with field applied either
parallel or perpendicular to the sample length, is used  to
determine the magnetoelastic coupling coefficient $$b^{\gamma,2}= -
\frac{2}{3} \frac{E_f}{1+\nu_f} \ \lambda$$ by the theory of
Lacheisserie and Peuzin \cite{lacheisserie-jmmm136} :
$$b^{\gamma,2}=b(H_{\parallel})- b(H_{\bot}) \mbox{ at } H_{\parallel} \mbox{
and } H_{\bot}> H_{sat} \mbox{ with }$$

\begin{equation}
\label{bgamma} b(H_i)=\frac{E_s}{6(1+\nu_s)}\frac{\theta_{f}(H_i) \,
t_s^2}{L \, t_f} \quad \quad  (i=\parallel,\bot)
\end{equation}

The $\parallel$ and $\bot$ indexes refer to the directions of
applied field: parallel or perpendicular to the sample length, which
in our case coincides with  the magnetic easy axis of the films (see
figure \ref{dispo}). $\lambda$ is the magnetostriction coefficient
of the film, $t_s$ is the substrate thickness, $t_f$ is the film
thickness, where $t_s \gg t_f$.
 $E_s$, $E_f$ and $\nu_s$, $\nu_f$, are, respectively, the
Young modulus and  Poisson ratio of the substrate (s) and of the
film (f).

For Corning glass substrates $E_s=60$ GPa and $\nu_s=0.27$.
Unfortunately the elastic parameters of the film cannot be
accurately determined and all techniques of magnetostriction
measurements of thin films provide only $b^{\gamma,2}$, with
$\lambda$ known only approximatively.

It is important to note that the  equation above is valid only  when
the maximum vertical sample deflection is very small : $L \theta \ll
t_s$, that is fulfilled in our experiment. In  this condition the
total curvature of the sample $R$ is an integral of local
deformations and depends neither on its shape nor on its
inhomogeneity. In this case, one can easily consider our real
situation, where the principal axis of the curvature (which
corresponds to the direction of sample magnetization) deviates from
the sample axis. The sample has, thus, both flexion and torsion. In
order to evaluate them, one can imagine a small narrow rectangle cut
from a cylindrical surface of radius $R$ at an angle $\varphi$ from
its axis. The rotation of the normal to the surface from one to the
other end of the rectangle is $\theta_n=L/R \ \sin \varphi$. Its
component in the direction of the sample axis is the flexion angle
$\theta_f=\theta_n \sin \varphi= L/R \ \sin^2 \varphi$, and its
transversal component is the torsion angle $\theta_t=\theta_n \cos
\varphi= \frac{1}{2} L/R \ \sin 2 \varphi$.

When large sample deformations are required for application, the
above conditions are not fulfilled and a much more complicated
theory has to be used, see
\cite{wetherhold_jap},\cite{wetherhold_apl}. In general, the elastic
properties of thin plates are described by non-linear equations
\cite{LL-elast}.


Another point to outline is that the correct definition of the
positive and negative flexion angle is important for the
determination of the signs of $b^{\gamma,2}$ and $\lambda$. The
flexion is positive when the sample turns towards the surface
covered by the magnetostrictive film (see Fig. \ref{dispo}). All our
measurements with TbCo/CoFe and TbFe/CoFe films give negative values
for $b^{\gamma,2}$ corresponding to positive $\lambda$ (elongation
of the material along the applied field) as reported in the
literature also for bulk TbFe and TbCo samples .


Domain observations were performed using the magneto-optical
longitudinal Kerr effect (MOKE microscopy \cite{Hubert-book}). The
microscope is of the split-path type, with an incidence angle of
25$^\circ$. The lenses, with a low numerical aperture ($\sim 0.1$),
allow for wide field imaging of the inclined sample with good
polarisation quality. The objective lens is fitted with a rotatable
wave plate ($\lambda / 20$) followed by a rotatable analyser, for
optimal contrast adjustment. The light source is a mercury lamp with
a pass-band filter around $\lambda= 546$ nm. A set of coils provides
field in the 3 directions. The incidence plane is parallel to the
vertical side of all the images of domains presented here.

\section{Results and discussion}

\begin{figure}[h]
\includegraphics[angle=0,width=8.5cm]{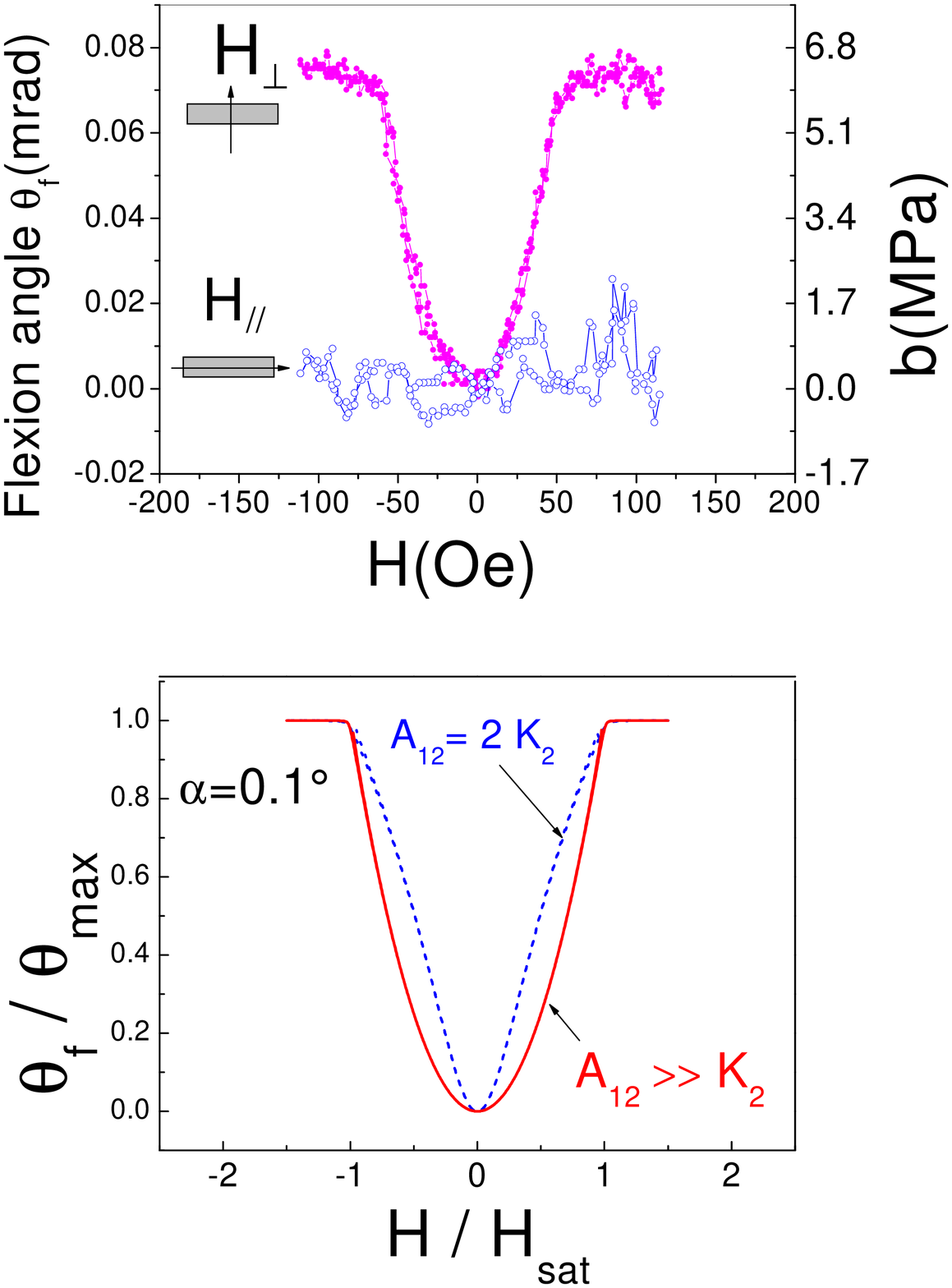}
\caption{\label{flexion} Flexion cycles for the field applied
parallel $H_{\parallel}$ and perpendicular $H_{\bot}$ to the easy
axis. Upper panel: experimental results. Lower panel: calculated
torsion cycles for two cases:  strongly coupled magnetizations of
magnetic layers ($A_{12}\gg 2 K_2 t_2$, $H_{sat}=2 K_2/\mu_0 M_1
t_1$), and moderate exchange between $M_1$ and $M_2$ ($A_{12}=4K_2
t_2$, $H_{sat}\approx 4.0 K_2 t_2 /\mu_0 M_1 t_1$)}
\end{figure}

The measurements of $b^{\gamma,2}$ of our multilayers are presented
in Figure \ref{flexion}. The fact that under the field applied along
the easy direction $\theta_f$ remains almost constant (zero) proves
the good alignment of the field.

We found a negative value of $b^{\gamma,2} \approx - 7$ MPa. One
should note that we have obtained an actuation field as low as 50 Oe
(4 kA/m) that gives a field sensitivity  much higher that  reported
by Betz for similar sample compositions and thicknesses
\cite{these-betz}. Our lower $b^{\gamma,2}$ is related to a lower
value of Curie temperature of our samples which should be close
enough to the room temperature in order to achieve maximal
magnetoelastic susceptibility.

\begin{figure}[h]
\includegraphics[angle=-90,width=8.5cm]{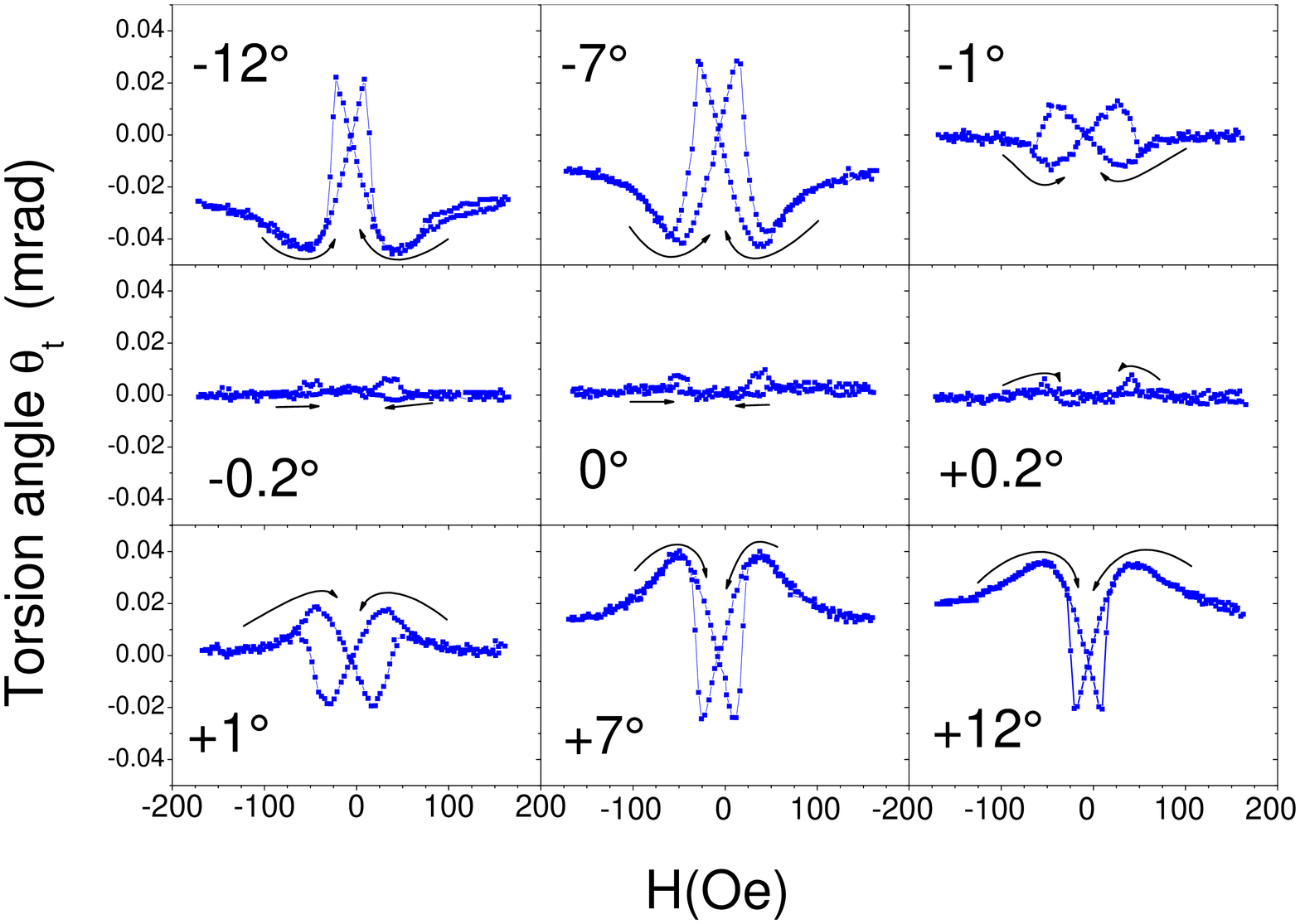}
\caption{\label{torsion}  Experimental torsion cycles for various
 angles $\alpha$ between hard axis and applied field direction. }
\end{figure}

 The flexion $\theta_f(H)$ cycle  does not evolve
much when the angle $\alpha$ between the hard axis and the external
magnetic field is varied a few degrees around zero. Contrarily, the
torsional angle loop $\theta_t(H)$ evolves considerably in a quite
unexpected  way as shown on figure \ref{torsion}. Not only the
amplitude but also the shape of the torsional cycle change
significantly when $\alpha$ is varied around zero. The
characteristic "butterfly"-like figures at small angles ($\alpha =
\pm 1^\circ$ in Fig. \ref{torsion}) was already found in similar
samples in
\cite{LeGall_IEEE_2001} and the question of their explanation was
open. Having in mind a simple monotonic magnetization rotation, one
would expect a simple sinusoidal form $\theta_t(H) \sim
\sin(H/H_{sat})$, so that the real complicated shape indicates a
more complex magnetization process.

Our systematic study of the angular dependence shows that the cycles are
antisymmetric around $\alpha=0$ (compare, for example, two frames in
Fig. \ref{torsion} corresponding to $\alpha= +12^\circ$ and
$-12^\circ$): when $\alpha$ changes sign, all the figure
$\theta_t(H)$ is reflected around the horizontal axis. On each
branch of $\theta_t(H)$ corresponding to increasing or decreasing
field  wide local extrema are separated by a sharp asymmetric peak
of opposite direction. It should be noted that the larger $\alpha$
is, the sharper are the central peaks. When the field is close to
the hard axis ($\alpha=0$) the observed 3 torsion oscillations
practically disappear. A small remaining irregular signal can be
explained by a small dispersion of the orientation of the anisotropy
axis throughout the sample that will be discussed further.

\begin{figure}[h]
\includegraphics[angle=-90,width=8.5cm]{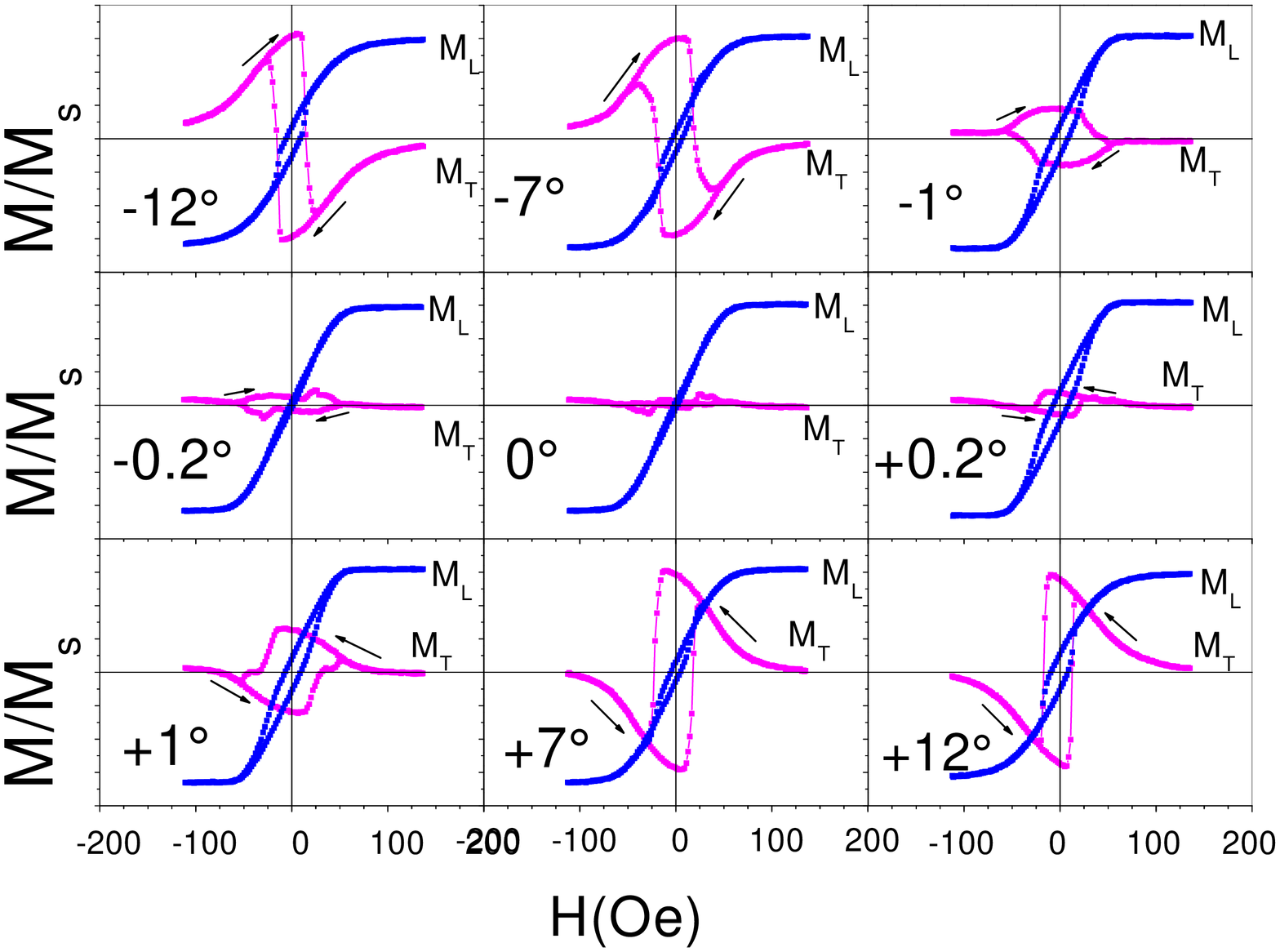}
\caption{\label{vsm} Experimental longitudinal ($M_L$) and
transverse ($M_T$) magnetization loops  for various angles $\alpha$
between hard axis and applied field direction. }
\end{figure}

A more transparent interpretation of the magnetization processes
behind the observed complicated magnetostrictive loops behaviour can
be obtained from  measurements of two components of the
magnetization. Figure \ref{vsm} presents the longitudinal (ordinary)
$M_L(H)$ and transverse magnetization $M_T(H)$ loops obtained for
different $\alpha$ angles.

The longitudinal magnetization  at $\alpha=0$ has nearly no
hysteresis as usual for the hard axis magnetization. At non-zero
$\alpha$ some hysteresis appears. We will show below that the
corresponding small coercive field is fully determined by the
magnetization rotation, the coercive field of domain wall motion
(corresponding to the field where the back and forth branches of the
hysteresis loop merge) being higher.

The hysteresis of $M_T(H)$ is much more visible. The transverse
loops are antisymmetric around $\alpha=0$ in a manner similar to the
torsional cycles: for positive $\alpha$ when magnetic field is
decreased the transverse magnetization $M_T$ increases first and
abruptly changes from positive to negative, whereas for negative
$\alpha$, $M_T$ first decreases and abruptly changes the sign in
opposite direction. It shows clearly that the monotonic
magnetization rotation, started from saturation, interrupts  at some
moment after the field inversion, and magnetization flips relatively
to the hard axis in order to be closer to the magnetic field again.
Then it continues to turn toward $\vec{H}$, now in the opposite
direction. The abrupt magnetic inversion corresponds to the sharp
peak on $\theta_t(H)$ (see Fig. \ref{torsion}).

\begin{figure}[h]
\includegraphics[angle=0,width=5cm]{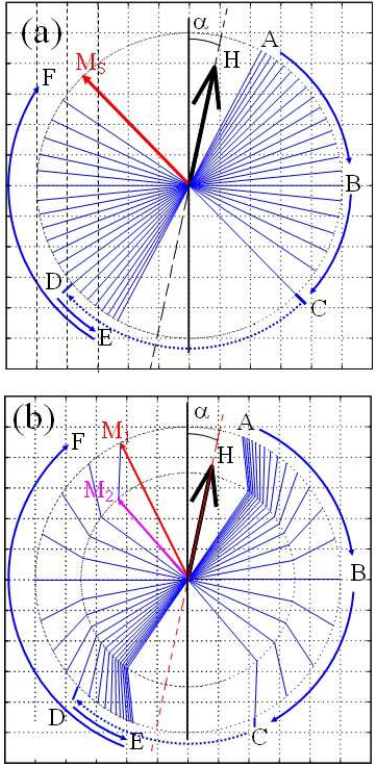}
\caption{\label{polar} Vectorial diagram of the magnetization
rotation in (TbFe/CoFe) multilayers. Direction of the applied field
$H$ (dashed line) is deviated from hard axis by $\alpha=12^\circ$.
In the figure easy axis is horizontal and hard axis is vertical.
(a)Approximation by a single magnetization vector $\vec{M}_s$
(Stoner-Wohlfarth). The fan of thin lines shows the sequence of the
rotation of $\vec{M}_s$ for field varying between $H_{max}=
\frac{3}{2} H_K$ and $-H_{max}$ with step $\Delta H =\pm 0.1 H_K$
($H_K=\frac{2K}{\mu_0 M_s}$). (b)Approximation by two coupled
magnetization vectors $\vec{M}_1$ and $\vec{M}_2$ corresponding to
magnetization of CoFe and TbCo layers (exchange between layers
$A_{12}= 4 K_2 t_2$).The sequence of the rotation is represented by
canted lines with the point on external circle corresponding to
$\vec{M}_1$ and the point on internal circle corresponding to
$\vec{M}_2$. Here $H_{max} = 3H_K (H_K= \frac{2 K_2 t_2}{\mu_0 M_1
t_1}$) and $\Delta H = \pm 0.2 H_K$. Points A,B,..F mark
characteristic moments of the magnetization rotation (see text). The
vector position between points D and E superimpose for the
descending and ascending field.}
\end{figure}

\begin{figure}
\includegraphics[angle=0,width=8.5cm]{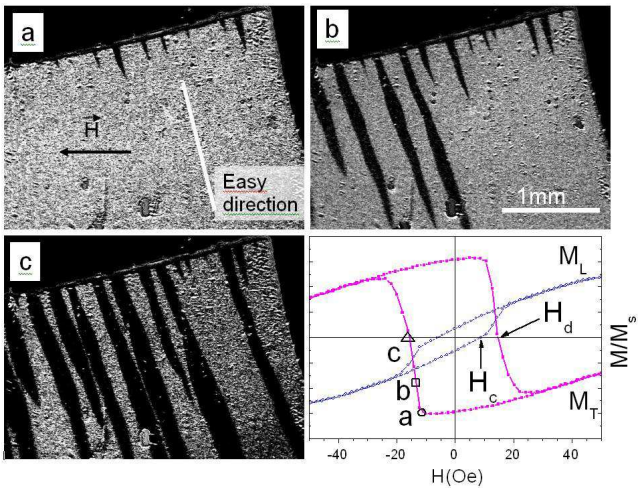}
\caption{\label{moke-12} MOKE microscopy images for magnetic field
deviating by  12$^\circ$ from the hard axis.  (a) $H=-10.5$ Oe (845
A/m). (b) $H=-12.9$ Oe (1030 A/m). (c) $H=-14.6$ Oe (1170 A/m) in
sequence after saturation in positive field. These points are
indicated on the corresponding measured magnetization loop $M_T(H)$
(central part) shown in the lower-right panel together with the
"standard" $M_L(H)$. The illustrated magnetization inversion
corresponds to the jump $C\rightarrow D$ in Fig. \ref{polar}
realized by  motion of domains. }
\end{figure}

We now explain the details of the magnetization rotation detected by
the vectorial measurements of $M_L(H)$ and $M_T(H)$ for the case
$\alpha=12^\circ$ (Fig. \ref{vsm}, see also the inset Fig.
\ref{moke-12}) using the  schematic  diagram of Fig.
\ref{polar}-(a). This behavior was  first described by Stoner and
Wohlfarth \cite{SW}.

Let us start from the strong field $H > H_K$ which is applied almost
perpendicularly to the easy axis and is large enough to approach the
longitudinal magnetization saturation: $ M_L \approx + M_s $ and
$M_T \approx 0$ (point A in Fig. \ref{polar}-(a)). When the field
starts to decrease, magnetization, evidently, rotates toward the
closest easy direction. $M_L(H)$ decreases and $M_T(H)$ increases
continuously as can be seen (Fig. \ref{vsm}, $\alpha=12^\circ$). For
$H=0$ the magnetization is aligned along the easy axis and $M_T
\approx + M_s$ (point B). When the field is reversed, at some moment
(point C) competition between Zeeman energy and the energy of the
uniaxial magnetic anisotropy causes an abrupt magnetization switch
to the opposite easy direction (point D) which becomes now closer to
the field direction ($M_T$ changes sign). This switch is clearly
visible on $M_T(H)$ while $M_L(H)$ only shows a small discontinuity
which is even not visible for $\alpha=7^\circ$ (see Fig. \ref{vsm}.)
After this transition the magnetization continues to rotate, {\it
now in the opposite direction},  till the negative saturation: $M_L
\approx -M_s$ and $M_T \approx 0$ again (point E). The return
magnetization branch is similar but it passes "by the other side"
with $M_T\approx -M_s$ at $H=0$ (from E to F and further).

In order to describe our multilayer system more precisely we
consider  CoFe and TbCo layers as two interacting magnetization
vectors, $\vec{M}_{1}$ and $\vec{M}_{2}$. We reduce the total energy
(per surface unit) to three terms only: Zeeman energy of the CoFe
layers, uniaxial magnetic anisotropy of TbCo layers and the
interlayer exchange energy:

\begin{eqnarray}
\label{2films}
 E(\phi_{1},\phi_{2})= &-& \mu_0 M_{1} t_1 H  \cos(\phi_{1}-\alpha)-
 K_{2} t_2 \sin^2(\phi_{2}) \nonumber \\
&-& A_{12}\cos(\phi_{1}-\phi_{2})
\end{eqnarray}

We neglect the Zeeman energy of the TbCo layers, because their
magnetization is much smaller than that of CoFe layers, and the
magnetic anisotropy energy of CoFe layers that are known as
magnetically very soft. The orientations of the magnetization of
both CoFe film $\phi_1$ and TbCo film $\phi_2$   are measured from
the hard axis direction (see Fig. \ref{dispo}). $t_1$ and $t_2$ are
the total CoFe and TbCo thicknesses ($t_1+t_2=t_f$).
Magnetostrictive and elastic energies do not enter to this equation
because, in our isotropic case, they do not depend on the
orientation of magnetization and  the curvature of the sample, due
to balance of these energies that is taken into account in equation
(\ref{bgamma}), 
remains constant. During the magnetization rotation considered here,
only the orientation of the curvature changes as it was described
above for the  explanation of the relation of torsion and flexion.

 The equilibrium angles $\phi_1^{\star}(H)$ and
$\phi_2^{\star}(H)$, which correspond to local minima of
$E(\phi_{1},\phi_{2})$ are numerically determined for successive
values of the external field H and for its various orientations
$\alpha$. Once the equilibrium angle is found, the longitudinal and
transverse magnetization are simply computed since they are given
principally by CoFe layers: $M_L = M_1 \cos(\phi_1^{\star}-\alpha)$
and $M_T= M_1 \sin(\phi_1^{\star}-\alpha)$. The flexion and torsion
angles are defined by the magnetostrictive layers TbCo:
$\theta_f=\theta_{max} \cos ^2(\phi_2^{\star})$ and
$\theta_t=\frac{1}{2} \theta_{max}\sin(2 \phi_2^{\star})$
correspondingly (see the explanations above).
Here

$$\theta_{max} =\frac{6(1+\nu_s)}{E_s}\frac{ L \, t_f}{t_s^2} b^{\gamma,2}$$
(see Eq. \ref{bgamma}).

For sufficiently strong exchange between the layers, $A_{12} \gtrsim
2 K_2 t_2$, the vectors $\vec{M}_1$ and $\vec{M}_2$ turn together,
the "magnetostrictive" vector $\vec{M}_2$ following the leading
control vector $\vec{M}_1$ with a delay (see the results of the
numerical solution for $A_{12} =4 K_2 t_2$ in Fig. \ref{polar}(b)).
This behavior is very similar to the rotation of a single
magnetization vector qualitatively described above for the classical
Stoner-Wolfarth model of a simple uniaxial magnetic film (Fig.
\ref{polar}(a)). Both models give the characteristic points
A,B,...,F with the abrupt magnetization inversion from C to D. If
$A_{12} \lesssim 2K_2 t_2$, this magnetization jump may disappear
(depending on the value of $\alpha$).

It should be noted that CoFe and TbCo magnetizations are known to be
 coupled antiparallel (\cite{these-betz}). Nevertheless, since in our
model we neglect the Zeeman energy of the TbCo layers, and since
$\theta_f=\theta_{max} \cos ^2(\phi_2^{\star})$ and
$\theta_t=\frac{1}{2} \theta_{max}\sin(2 \phi_2^{\star})$ the
magnetization, torsion and flexion loops are not sensitive to the
sign of $A_{12}$. Taking this into account, we have  presented the
case of  positive $A_{12}$ (parallel coupling) for clarity of figure
\ref{polar}.

 In our experimental results we do not see any effect of the
delay between $\vec{M}_1$ and $\vec{M}_2$. For example, we reproduce
the U-like shape of the measured flexion curve $\theta_f(H)$ only
for $A_{12} \gg K_2 t_2$
 while  a moderate $A_{12}$ gives a
remarkably different triangular shape (Fig. \ref{flexion}: down
panel). So, our practical multilayers can be well described by the
simplest model ($A_{12} = \infty$), usually called the
Stoner-Wohlfarth model, where $\vec{M}_1$ and $\vec{M}_2$ are
parallel and can be, thus, represented by only  one magnetization
vector $\vec{M}_s$ (\cite{SW},\cite{Thiaville-SW}):

$$ \frac{E_{SW}(\phi)}{t_f}= - \mu_0 M_s H \cos(\phi-\alpha)-  K \sin^2(\phi)$$

The course of the magnetization rotation $\phi^*(H)$  described
above (Fig. \ref{polar}-(a)) was obtained by minimization of this
energy for varying field between $H_{max}=\frac{3}{2} H_k$ and
$-H_{max }$.

\begin{figure}[h]
\includegraphics[angle=-90,width=8.5cm]{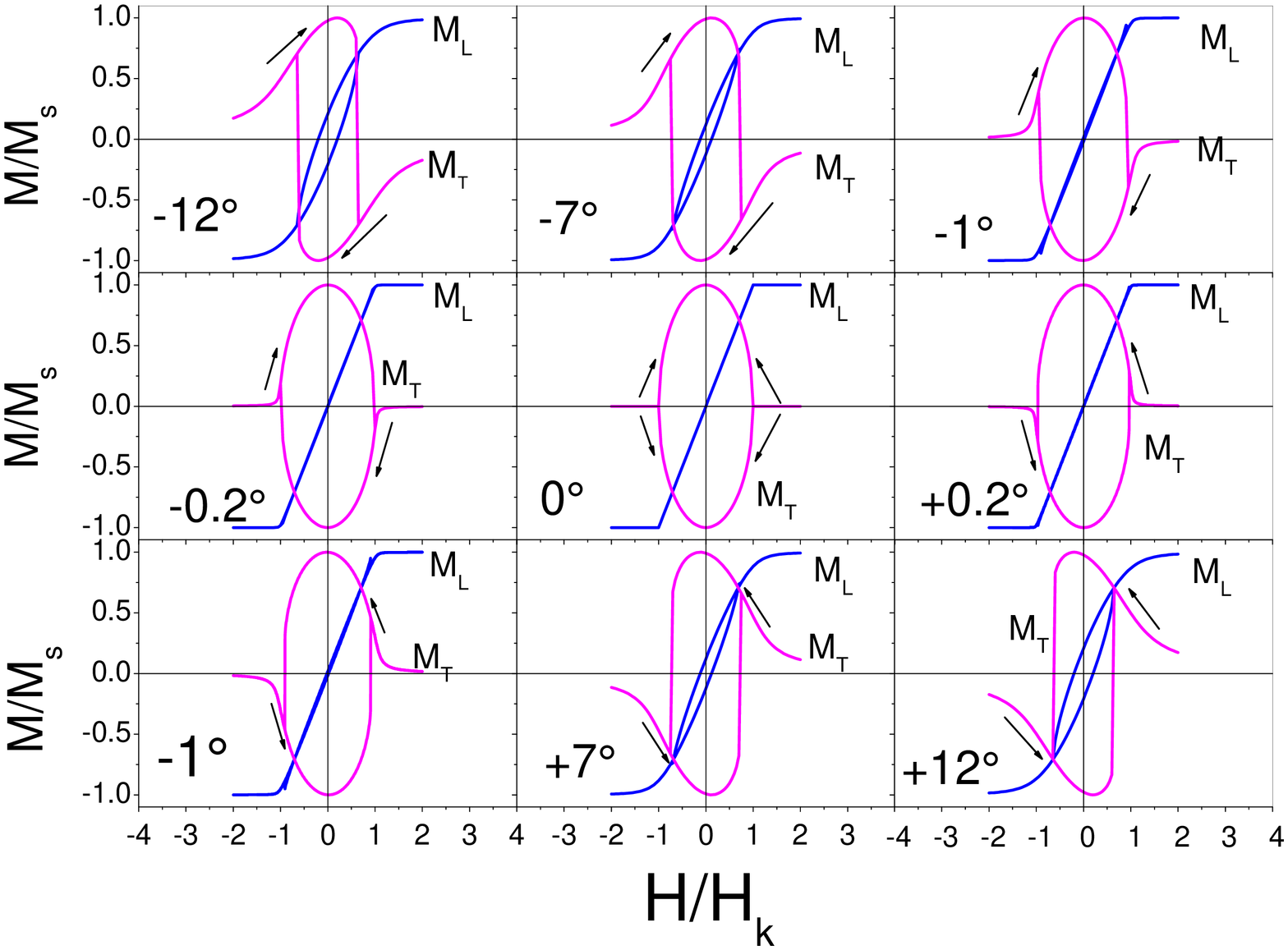}
\caption{\label{simul-aim}  Longitudinal ($M_L$) and transverse
($M_T$) magnetization loops for various angles $\alpha$ between hard
axis and applied field direction  calculated in the framework of the
model of the single magnetization vector.}
\end{figure}

\begin{figure}
\includegraphics[angle=-90,width=8.5cm]{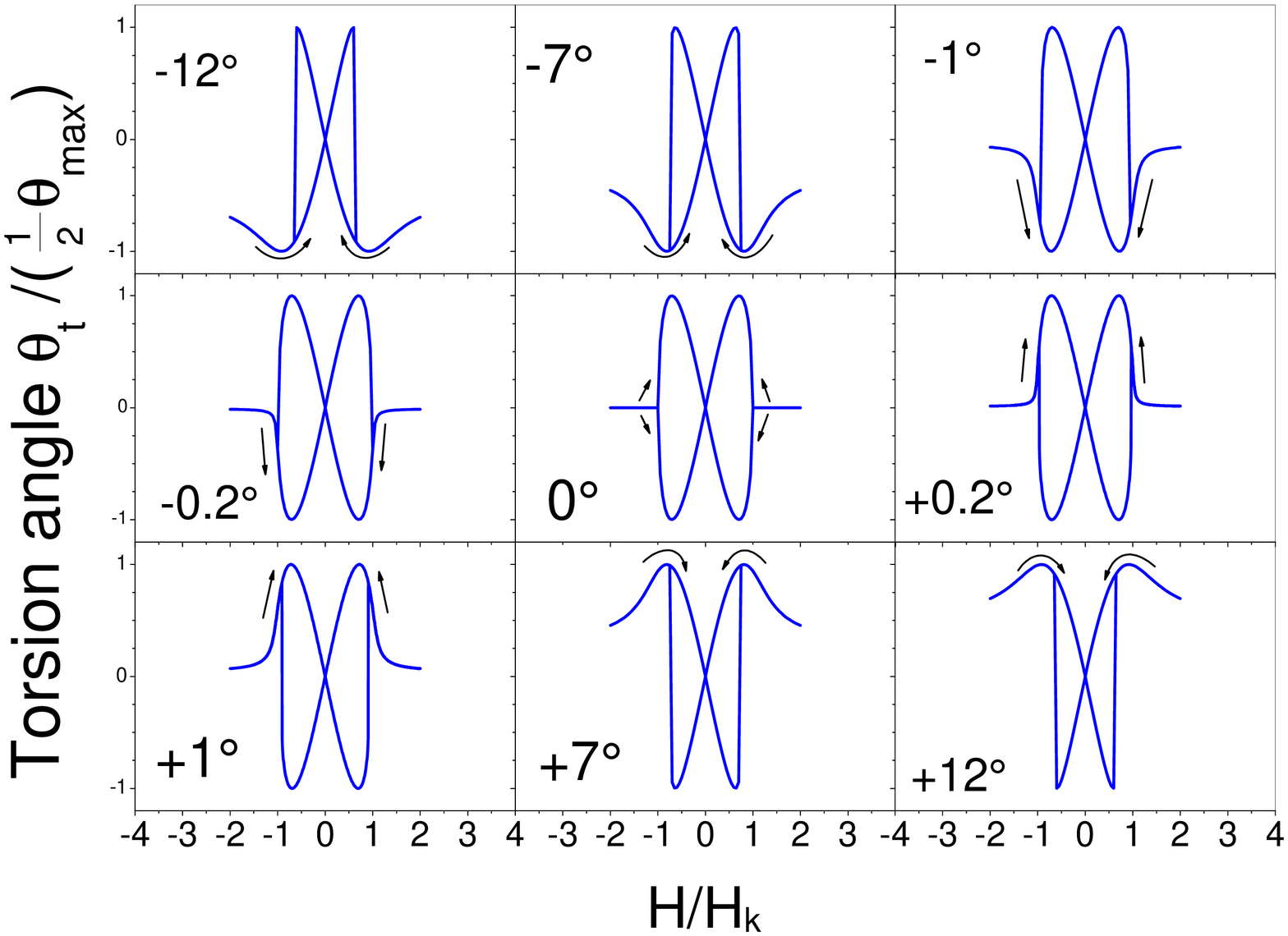}
\caption{\label{simul-torsion} Torsion cycles  magnetization loops
for various angles ($\alpha$) between hard axis and applied field
direction corresponding to the magnetization loops given in Fig.
\ref{simul-aim}.}
\end{figure}

Figures of simulations of magnetization (Fig. \ref{simul-aim}) and
magnetostriction (Fig. \ref{simul-torsion}) loops are obtained from
this solution using $M_{L,T}(\phi_1^\star)$ and
$\theta_{f,t}(\phi_2^\star)$ given above with $\phi_1^* = \phi_2^* =
\phi^*$ and $M_1=t_f /t_1 \ M_s$.

General features observed experimentally are well reproduced by this
simple model. This is particularly clear for  hard axis longitudinal
magnetization $M_L(H)$ (Fig. \ref{vsm} and Fig. \ref{simul-aim},
$\alpha = 0$). It is linear until the saturation as for the
classical case of monophase samples with good uniaxial anisotropy.
In this model the anisotropy field $H_K$ where the extrapolation of
the linear part intercepts the saturation magnetization is usually
introduced. For our sample $H_K = \frac{2K}{\mu_0 M_s} = \frac{2 K_2
t_2}{\mu_0 M_1 t_1}= 50$ Oe (4 kA/m).

When $\alpha$ deviates from zero the longitudinal loops $M_L(H)$ are
no longer linear and open up more and more as seen experimentally.
The characteristic "strange" butterfly shape of the torsion cycle
$\theta_t(H)$ is also well reproduced by this model for angles
$\alpha$ not too close to zero without necessity to consider more
complex effects as was suggested in \cite{Legall-Jap}. The
experimentally observed  inversion of the "wings" for opposite
$\alpha$ follows naturally from the model as well.

When  $H$ is parallel to the hard axis, two possible directions of
the magnetization rotation ( clockwise or counter clockwise) and two
corresponding branches of the transverse magnetization $M_T(H)$
(positive  and negative arcs) are equivalent (Fig. \ref{simul-aim},
$\alpha=0$).  The corresponding nearly zero experimental values for
$M_T(H)$ (Fig. \ref{vsm}, $\alpha=0$) suggest that the sample is
subdivided into magnetic domains where both  opposite possibilities
are realized. The magnetization in the domains alternates in a way
that on average the total transverse magnetization compensates.

As soon as $\alpha$ differs from zero, one branch is favorable for
the decreasing field and the opposite for the increasing field so
that the experimental $M_T(H)$ is no longer compensated.

This remark is also valid for the torsion angle loops $\theta_t(H)$
(Figs. \ref{torsion} and \ref{simul-torsion}). For $\alpha=0$ the
calculated back and forth loops are equivalent showing one minimum
and one maximum. The experimental $\theta_t(H)$ is compensated in
the same way as $M_T(H)$. For experimental measurements at $\alpha
\ne 0$, a "butterfly" loop opens up similarly  to the calculated
figures.  At this moment the third additional extremum corresponding
to the abrupt magnetization inversion starts to be clearly visible.

Obviously, the above model of coherent rotation (both in the
simplified case of the single $\vec{M}_s$ vector and in the case two
vectors $\vec{M}_1$ and $\vec{M}_2$) does predict  the observed
abrupt magnetization inversion. Nevertheless, in the experiment the
transition between two magnetization "branches" is not really
abrupt:  for example, at $\alpha=\pm 12^{\circ}$ the transition
occurs over a field range of $\Delta H \approx 10$ Oe (800 A/m). It
is well known that the abrupt coherent magnetization rotation
predicted by the Stoner-Wolfarth model, at the moment when the
metastable solution of $\phi^*$ disappears, can be realized only in
small samples. In large samples the inversion from the metastable to
the stable orientation is realized before reaching the instability
field by means of the nucleation and motion of magnetic domains. So,
in order to understand all details of the observed phenomena, one
has to understand the peculiarities of the domain formation and
evolution in this case.

\begin{figure}
\includegraphics[angle=0,width=8.5cm]{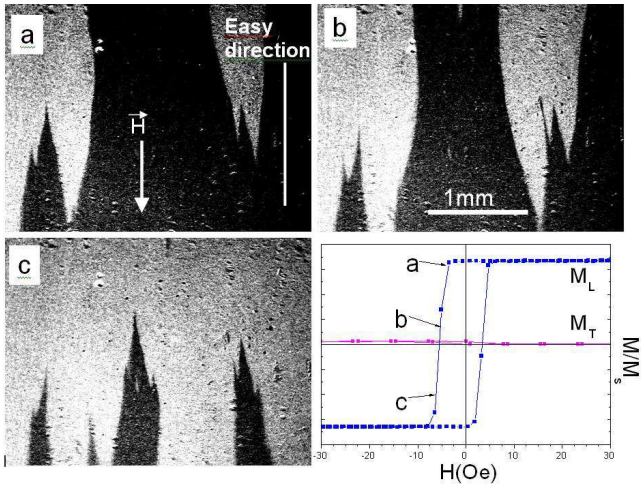}
\caption{\label{moke-para} MOKE microscopy images for magnetic field
($H_{easy}$) applied along the easy axis following positive
saturation. (a): first appearance of domains $H_{easy}=$ -4.1 Oe
(325 A/m), (b): $H_{easy}=$ -4.4 Oe (350 A/m),(c): $H_{easy}=$ -4.8
Oe (380 A/m). The corresponding magnetization loop is shown in the
lower-right corner. The moments where the Kerr images were obtained
are marked by arrows (from a to c).}
\end{figure}

We have directly observed the domains in our experimental situation
by means of the longitudinal magneto-optical Kerr effect.
The Kerr images obtained for field applied parallel to the easy axis
($\alpha=90 ^\circ$) are presented in Fig. \ref{moke-para}. The
incidence plane is also parallel to this axis. In this geometry, the
observed magneto-optical contrast  is proportional to the projection
of $\vec{M}$ onto the easy axis.
Here, and in all figures below, the image size is 3 mm $\times$ 2.2
mm. Frames a-c in Fig. \ref{moke-para}, showing successive images
obtained in decreasing field, correspond to the descending branch of
the magnetization after the saturation in the positive field $H>+10$
Oe (800 A/m). At the positive saturation, all the image is dark
(magnetization is directed upward). The opposite bright spontaneous
magnetic domains appear at the sample edge where the stray field is
maximal (at the top of the image) only when the field reverses and
reaches $\sim  -4$ Oe (-320 A/m). Frame (a) corresponds to the
moment after the first appearance of the domains. When field sweep
continues, the bright domains grow and expel the dark ones (frames
(b) and (c)). The last dark domains are seen for $H= -5.1$ Oe (405
A/m).  This observation corresponds well to the magnetization loop
measured for $\alpha=90^\circ$ (see inset in Fig. \ref{moke-para})
and its rectangular shape is usual when the field is applied along
the easy magnetic direction. In this case spontaneous domains occur
only in a narrow field range ($\sim$ 1 Oe (80 A/m)) where they
provide the magnetization inversion. It should be noted that the
domains continue to move even if the field is kept constant; this
thermally activated motion gives some modification of the hysteresis
loop width as function of the rate of the applied field sweep.

\begin{figure}
\includegraphics[angle=0,width=8.5cm]{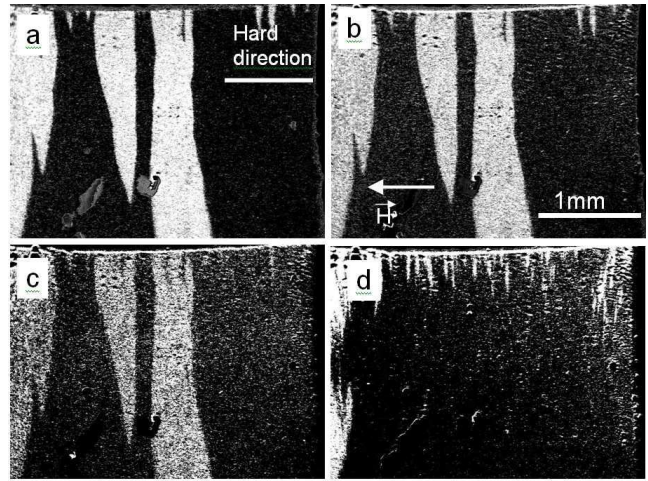}
\caption{\label{moke-perp} MOKE microscopy images for magnetic field
applied along the hard direction. (a) $H=0$, the domains are
prepared by an easy axis magnetization sequence (see text).
Subsequent fields applied along the hard direction: (b)
$H_{hard}=$30.6 Oe (2450 A/m) , (c) $H_{hard}=$44.2 Oe (3540 A/m),
(d) $H=0$ after saturation at $H_{hard} > 50$ Oe (4 kA/m) along the
hard axis. The sample edge is visible at the top of these images.}
\end{figure}

Now we pass from the classical observation of the domain motion
under  the field applied along the easy axis to our experiments on
the hard axis magnetization. In the images below (Fig.
\ref{moke-perp}) the field is applied horizontally while the
incidence plane is kept in the same position (vertical). The
magneto-optical contrast is proportional to the projection of
$\vec{M}$ onto the vertical axis as before.

We start from the domain configuration created during the previous
experiment with $\vec{H} \parallel$ easy axis (see Fig.
\ref{moke-para}). The initial domain configuration shown in Fig.
\ref{moke-perp}-a is obtained after the following sequence with
field applied along the easy axis: negative saturation, field
inversion to $H=+4$ Oe (320 A/m) (formation of domains), $H=0$
(domain structure is frozen in). The shape of these domains does not
vary when the magnetic field is applied perpendicular to the easy
axis ($\alpha=0$ Fig. \ref{moke-perp} b-c). Only the contrast
between opposite domains disappears progressively corresponding to
the monotonous rotation of the magnetization inside domains towards
the field. The field at which the contrast becomes zero corresponds
to the saturation field of the hard axis magnetization $H_{sat}=50$
Oe (4 kA/m) obtained in VSM measurements (figure \ref{vsm} for
$\alpha=0$).

%

At the saturation ($H>H_{sat}$), the image intensity is uniform
(average between the dark and bright values of the initial domains).
Upon the following reduction of the perpendicular field below
$H_{sat}$ the initial artificially created domains never appear
again. Only some very small spontaneous domains nucleate at the
sample edges just below $H_{sat}$. They remain immobile until the
opposite saturation just as discussed above in the case of the
"artificial" domains. Their contrast reaches a maximum at $H=0$
(Fig.\ref{moke-perp}-d). It is interesting to  note  that the major
area  of the shown part of the sample became dark, i.e. the
magnetization has chosen nearly everywhere here the upward easy
direction. If it would be so in the whole sample, we should have
observed the nice large "butterflies" for $\theta_t(H)$ and arcs for
$M_T$ which we calculated. The observed compensation of these
signals means that the nearly homogeneous magnetization shown in
Fig.\ref{moke-perp}-d at one sample end does not extend towards its
other end.

\begin{figure}
\includegraphics[angle=-90,width=8.5cm]{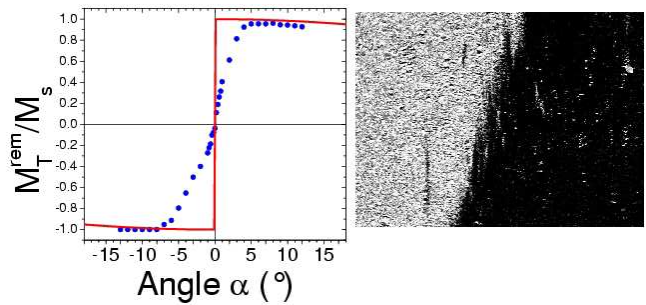}
\caption{\label{mtr} Characterisation of the inhomogeneity of the
sample magnetic anisotropy. Left panel : dependance of the
transverse remanent magnetization $M_T^{rem}$ of the sample on the
field orientation $\alpha$. Solid circles represent the experimental
masurements and solid line shows $M_T^{rem}(\alpha)$ expected for an
ideal sample with a well defined easy axis. Right panel: MOKE image
of the macrodomain boundary appeared in a central part of the sample
at $H=0$ after the hard axis saturation. The magnetization here is
vertical: downward in the left bright macrodomain and upward in the
right dark macrodomain.}
\end{figure}

Because of an inhomogeneity of the orientation of the easy axis the
different parts of the sample will have opposite orientations of the
magnetization rotation towards the closest easy direction. We indeed
observe a  magnetic separation boundary in the central part of the
sample (Fig. \ref{mtr}). This boundary has a complex fine structure
analogous to the structure of the spontaneous domains at the sample
edges and also remains immobile under the hard axis field. At zero
field its magneto-optical contrast is maximal because the
macrodomains separated by it have opposite magnetization along the
easy axis. It should be noted that this is not a normal $180^\circ$
domain wall. The usual $180^\circ$ domain wall is nearly parallel to
the easy axis and moves across the sample by the magnetic field. On
the contrary, the position of the macrodomains boundary is fixed on
the line where the local easy axis is parallel to the applied field.
More precisely, this line is the solution of the equation
$\alpha_0(x,y)= \alpha$  where $\alpha_0$ is the local orientation
of the easy axis at point $x,y$ on the sample surface.

It is obvious that with a different orientation of the applied
magnetic field  $\alpha$, this boundary appears at a different
position. In order to characterize this effect  we have plotted the
remanent {\it transverse} magnetization $M_T^{rem}$
($M_T^{rem}=M_T(H=0$)) as a function of the angle $\alpha$ between
the hard axis direction and the magnetic field (see Fig.\ref{mtr}).
For $\alpha$=0, in a homogeneous sample, one should observe
$M_T^{rem} = \pm M_s$ since both rotation branches are equivalent,
as discussed previously. When $\alpha$ departs from zero $\mid
M_T^{rem}\mid $ decreases and reaches zero for $\alpha=\pm 90
^\circ$ when the field is applied along the easy axis:
$$ M_T^{rem} = M_s \, \mbox{sign}(\alpha) \, \cos(\alpha). $$
In the reality, the central step at $\alpha=0$ is smeared according
to the distribution $\alpha_0(x,y)$  and the corresponding gradual
displacement of the boundary between two opposite macro-domains with
$M_T^{rem}=\pm M_s$. In our case the experimental values match the
model  for $|\alpha | \gtrsim 6 -7^\circ$. This allows to quantify
the easy axis distribution within the sample: its maximum deviation
from the average position is about $\pm 6^\circ$.  When the magnetic
field is applied in this interval around the hard axis direction the
sample is divided onto correspondingly proportioned opposite
macro-domains, while when $|\alpha| > 6^\circ$ the sample is
monodomain. In our sample the observed structure of the transverse
remanent magnetization corresponds to a gradual monotonous rotation
of the easy axis across the sample. So, one can expect that in a
smaller sample this angular range where the macro-domains appear
would be narrower. It is evident, that each macro-domain represents
the  solutions for $M_T(H)$ and $\theta_t(H)$ shown above at
positive or negative small values of $\alpha$, which have the same
amplitude and shape, but change sign from one domain to another (see
central frames of Figs. \ref{simul-aim} and \ref{simul-torsion}).
So, the total value averaged over both domains will be compensated
at $\alpha \approx 0$.
%

In order to illustrate the "ideal" process of the sample
magnetization at $|\alpha| > 6^\circ$,  when the magnetization
rotation is homogeneous over the whole sample, we present Kerr
images obtained at a field deviating by $\sim 12 ^\circ$ from the
hard axis direction (Fig. \ref{moke-12}). We see how, after the
initial saturation magnetization in the negative field, the
magnetization inversion starts in a positive field by nucleation of
spontaneous magnetic domains at the sample edge where the stray
field is maximum (Fig. \ref{moke-12}-a). The motion of domain walls
 extends over a field range of some {\OE}rsted (Fig.
\ref{moke-12}-b,c) which is considerably larger than in the case of
the easy axis magnetization presented above ($\sim 10$ Oe = 800 A/m
contrary to $\sim 1$ Oe =80 A/m) but remains rather narrow in
comparison with the overall width of the hysteresis ($H_d \sim 30$
Oe = 2400 A/m) as seen in the hysteresis loops presented in the
inset.

It is important to note here the difference between the transverse
hysteresis width ($H_d$, the field at which the magnetization vector
flips from one side to the other, i.e. $M_T(H_d)=0$) and the usual
coercive field ($H_c$, where $M_L(H_c)=0$). $H_d(\alpha)$ represents
the coercive field of the domain wall motion reached when the
corresponding force is equilibrated by the difference of the Zeeman
energy density : $F_C(\vec{M}_{+},\vec{M}_{-}) = \vec{H_d} \cdot
(\vec{M}_{+} - \vec{M}_{-})$. Here $\vec{M}_{+}$, $\vec{M}_{-}$ are
magnetizations of domains with opposite transversal components.

For $\alpha \sim 0$, $H_d$ becomes as large as the saturation field
$H_K$ and the domains remain immobile between $\pm H_d$ as
demonstrated  in Fig. \ref{moke-perp}. $H_d \equiv H_c$ only for
large deviation from the hard axis. For sufficiently small $\alpha$
the  magnetization "jump" on  the magnetization hysteresis curve
$M_L(H)$ (the short interval of the domain motion) appears
\emph{after}  it crosses the horizontal axis. So, $H_c$ has nothing
to do with the domain wall motion and does not correspond to any
discontinuity of the magnetization process. In this case (for
example for $\alpha=12^\circ$, Fig. \ref{moke-12}), $H_c$ represents
the hysteresis of the homogeneous magnetization rotation. Obviously
for $\alpha=0$, $H_c=0$.

It is interesting that the coherent magnetization rotation dominates
the major part of the hysteresis even for so big samples, where
normally the magnetization by domains is considered as the most
important. These features were already observed in the case of
 simple monolayer films by Prutton
\cite{Prutton} and in spin-valve multilayers \cite{Labrune}. The
model of the coherent magnetization rotation explains the
experiments so well because, when the pinning of the domain walls is
sufficiently strong, or the domain wall nucleation sufficiently
difficult, the domains appear and move just before the moment of the
rotational instability. The resulting behavior does not differ very
much from the calculated abrupt magnetization inversion.

Very few investigations of domains in magnetostrictive multilayers
are known. Chopra and co-workers \cite{Chop-99} showed that in their
demagnetized state individual layer of as deposited TbFe/CoFe are
single domain due to stray field coupling between adjacent layers.
Our measurements and direct observations, on the contrary, show that
all layers are connected together by magnetic exchange and the
magnetization in the whole stack of the layers turns as a single
magnetic vector. The magnetic domains which inevitably appear in our
sample are the same in all layers (the magneto-optical contrast is
always homogeneous) and the domain walls go through the whole film.

Recently Shull and coworkers  \cite{MS-MOIF} studied the influence
of stress on domain wall structure in similar multilayers using the
magneto-optic indicator film (MOIF) technique that reveals domains
by an indirect effect of their stray field. They show that
magnetization reversibly rotates when stress is applied whereas
magnetization rotation under field application is irreversible and
includes wall motion. This work of Shull et al. is limited only to
easy axis magnetization when there is no magnetostrictive
deformation of the sample. We show that in the practically important
geometry of the hard axis magnetization, when the magnetostrictive
deformation is maximal, the magnetization rotation is dominating.
Our observations via the direct magneto-optical contrast give a more
clear image of domains and the magnetization distribution inside
them. Studies of the influence of the applied stresses on the domain
structure and motion of the domain walls in TbCo/CoFe multilayers
will be published elsewhere.

\section{Conclusion}

As result of the systematic angular measurements  of all components
of magnetization and magnetostrictive deformation together with a
simple model, the magnetization process of magnetostrictive
TbCo/CoFe multilayer has been fully understood. We have shown that
even in large (centimeter size) samples the major part of the
magnetic hysteresis is related to a coherent magnetization rotation
inside domains. The coherent magnetic rotation is broken only once
when a rather abrupt magnetization reversal occurs with appearance
of spontaneous magnetic domains. In addition, we have revealed the
appearance of stationary macro-domains related to the sample
inhomogeneity: They explain the disappearance of the torsion angle
loop and transverse magnetization when magnetic field is applied
along the hard axis. A small misalignment of magnetic field modifies
noticeably these loops as the ratio of opposite macro-domains
changes. The appearance of the micro- and macro-domains in this
system and the related effects should be taken into account in
further development of technical applications of the
magnetostrictive films. This is particularly important for the
reduction of the lateral size of the micro-actuators.

\begin{acknowledgments}

F.P. thanks Brest M\'etropole Oc\'eane for  financial support(PhD
fellowship).

We are grateful to the group of Ph. Pernod (IEMN-Lille) for the
collaboration (see Ref.
\cite{tiercelin-jmmm249},\cite{LeGall_IEEE_2001}), which initiated
this work. In particular, N. Tiercelin is acknowledged for the
original development of the measurements.

\end{acknowledgments}

\newpage
\bibliographystyle{apsrev}

\begin{thebibliography}{21}
\expandafter\ifx\csname
natexlab\endcsname\relax\def\natexlab#1{#1}\fi
\expandafter\ifx\csname bibnamefont\endcsname\relax
  \def\bibnamefont#1{#1}\fi
\expandafter\ifx\csname bibfnamefont\endcsname\relax
  \def\bibfnamefont#1{#1}\fi
\expandafter\ifx\csname citenamefont\endcsname\relax
  \def\citenamefont#1{#1}\fi
\expandafter\ifx\csname url\endcsname\relax
  \def\url#1{\texttt{#1}}\fi
\expandafter\ifx\csname urlprefix\endcsname\relax\def\urlprefix{URL
}\fi \providecommand{\bibinfo}[2]{#2}
\providecommand{\eprint}[2][]{\url{#2}}

\bibitem[{\citenamefont{Claeyssen et~al.}(1997)\citenamefont{Claeyssen,
  Lhermet, Letty, and Bouchilloux}}]{JAC-actuators}
\bibinfo{author}{\bibfnamefont{F.}~\bibnamefont{Claeyssen}},
  \bibinfo{author}{\bibfnamefont{N.}~\bibnamefont{Lhermet}},
  \bibinfo{author}{\bibfnamefont{R.~L.} \bibnamefont{Letty}}, \bibnamefont{and}
  \bibinfo{author}{\bibfnamefont{E.}~\bibnamefont{Bouchilloux}},
  \bibinfo{journal}{J. Alloys and Compounds.} \textbf{\bibinfo{volume}{258}},
  \bibinfo{pages}{61} (\bibinfo{year}{1997}).

\bibitem[{\citenamefont{Pasquale}(2003)}]{pasquale}
\bibinfo{author}{\bibfnamefont{M.}~\bibnamefont{Pasquale}},
  \bibinfo{journal}{Sensors and Actuators A} \textbf{\bibinfo{volume}{106}},
  \bibinfo{pages}{142} (\bibinfo{year}{2003}).

\bibitem[{\citenamefont{Quandt and Ludwig}(1999)}]{Quandt-jap85}
\bibinfo{author}{\bibfnamefont{E.}~\bibnamefont{Quandt}} \bibnamefont{and}
  \bibinfo{author}{\bibfnamefont{A.}~\bibnamefont{Ludwig}},
  \bibinfo{journal}{J. Appl. Phys.} \textbf{\bibinfo{volume}{85}},
  \bibinfo{pages}{6232} (\bibinfo{year}{1999}).

\bibitem[{\citenamefont{Quandt et~al.}(97)\citenamefont{Quandt, Ludwig, Betz,
  Mackay, and Givord}}]{Quandt-jap81}
\bibinfo{author}{\bibfnamefont{E.}~\bibnamefont{Quandt}},
  \bibinfo{author}{\bibfnamefont{A.}~\bibnamefont{Ludwig}},
  \bibinfo{author}{\bibfnamefont{J.}~\bibnamefont{Betz}},
  \bibinfo{author}{\bibfnamefont{K.}~\bibnamefont{Mackay}}, \bibnamefont{and}
  \bibinfo{author}{\bibfnamefont{D.}~\bibnamefont{Givord}},
  \bibinfo{journal}{J. Appl. Phys.} \textbf{\bibinfo{volume}{81}},
  \bibinfo{pages}{5420} (\bibinfo{year}{97}).

\bibitem[{\citenamefont{Tiercelin et~al.}(2002)\citenamefont{Tiercelin, {\relax
  Ben Youssef}, Preobrazhensky, Pernod, and {\relax Le
  Gall}}}]{tiercelin-jmmm249}
\bibinfo{author}{\bibfnamefont{N.}~\bibnamefont{Tiercelin}},
  \bibinfo{author}{\bibfnamefont{J.}~\bibnamefont{{\relax Ben Youssef}}},
  \bibinfo{author}{\bibfnamefont{V.}~\bibnamefont{Preobrazhensky}},
  \bibinfo{author}{\bibfnamefont{P.}~\bibnamefont{Pernod}}, \bibnamefont{and}
  \bibinfo{author}{\bibfnamefont{H.}~\bibnamefont{{\relax Le Gall}}},
  \bibinfo{journal}{J. Magn. Magn. Mater.} \textbf{\bibinfo{volume}{249}},
  \bibinfo{pages}{519} (\bibinfo{year}{2002}).

\bibitem[{\citenamefont{Landau and Lifchitz}(1987{\natexlab{a}})}]{LandauL}
\bibinfo{author}{\bibfnamefont{L.}~\bibnamefont{Landau}} \bibnamefont{and}
  \bibinfo{author}{\bibfnamefont{E.}~\bibnamefont{Lifchitz}},
  \emph{\bibinfo{title}{Course of Theoretical Physics}}, vol.
  \bibinfo{volume}{5(Statistical physics) \S 144 and vol. 8 (Electrodynamics of
  Continuous Media)\S 46} (\bibinfo{publisher}{MIR},
  \bibinfo{year}{1987}{\natexlab{a}}).

\bibitem[{\citenamefont{Masson et~al.}(2004)\citenamefont{Masson, Juraszek,
  Ducloux, Euphrasie, Jay, Spenato, Teillet, {\relax Le Gall}, Preobrazhensky,
  and Pernod}}]{Mossbauer-Neel2004}
\bibinfo{author}{\bibfnamefont{S.}~\bibnamefont{Masson}},
  \bibinfo{author}{\bibfnamefont{J.}~\bibnamefont{Juraszek}},
  \bibinfo{author}{\bibfnamefont{O.}~\bibnamefont{Ducloux}},
  \bibinfo{author}{\bibfnamefont{S.}~\bibnamefont{Euphrasie}},
  \bibinfo{author}{\bibfnamefont{{\relax J.-Ph}.}~\bibnamefont{Jay}},
  \bibinfo{author}{\bibfnamefont{D.}~\bibnamefont{Spenato}},
  \bibinfo{author}{\bibfnamefont{J.}~\bibnamefont{Teillet}},
  \bibinfo{author}{\bibfnamefont{H.}~\bibnamefont{{\relax Le Gall}}},
  \bibinfo{author}{\bibfnamefont{V.}~\bibnamefont{Preobrazhensky}},
  \bibnamefont{and} \bibinfo{author}{\bibfnamefont{P.}~\bibnamefont{Pernod}},
  in \emph{\bibinfo{booktitle}{9\`{e}me Colloque Louis N\'{e}el-Couches minces
  et nanostructures magn\'{e}tiques}} (\bibinfo{year}{2004}), pp.
  \bibinfo{pages}{poster V--42}.

\bibitem[{\citenamefont{du~Tr\'{e}molet~de Lacheisserie and
  Peuzin}(1994)}]{lacheisserie-jmmm136}
\bibinfo{author}{\bibfnamefont{E.}~\bibnamefont{du~Tr\'{e}molet~de
  Lacheisserie}} \bibnamefont{and} \bibinfo{author}{\bibfnamefont{J.~C.}
  \bibnamefont{Peuzin}}, \bibinfo{journal}{J. Magn. Magn. Mater.}
  \textbf{\bibinfo{volume}{136}}, \bibinfo{pages}{189} (\bibinfo{year}{1994}).

\bibitem[{\citenamefont{Guerrero and Wetherhold}(2003)}]{wetherhold_jap}
\bibinfo{author}{\bibfnamefont{V.~H.} \bibnamefont{Guerrero}} \bibnamefont{and}
  \bibinfo{author}{\bibfnamefont{R.~C.} \bibnamefont{Wetherhold}},
  \bibinfo{journal}{J. Appl. Phys.} \textbf{\bibinfo{volume}{97}},
  \bibinfo{pages}{6659} (\bibinfo{year}{2003}).

\bibitem[{\citenamefont{Wetherhold and Chopra}(2001)}]{wetherhold_apl}
\bibinfo{author}{\bibfnamefont{R.~C.} \bibnamefont{Wetherhold}}
  \bibnamefont{and} \bibinfo{author}{\bibfnamefont{H.}~\bibnamefont{Chopra}},
  \bibinfo{journal}{Appl. Phys. Lett.} \textbf{\bibinfo{volume}{79}},
  \bibinfo{pages}{3818} (\bibinfo{year}{2001}).

\bibitem[{\citenamefont{Landau and Lifchitz}(1987{\natexlab{b}})}]{LL-elast}
\bibinfo{author}{\bibfnamefont{L.}~\bibnamefont{Landau}} \bibnamefont{and}
  \bibinfo{author}{\bibfnamefont{E.}~\bibnamefont{Lifchitz}},
  \emph{\bibinfo{title}{Course of Theoretical Physics}}, vol.
  \bibinfo{volume}{7(Theory of elasticity) \S 14} (\bibinfo{publisher}{MIR},
  \bibinfo{year}{1987}{\natexlab{b}}).

\bibitem[{\citenamefont{Hubert and Sch\"{a}fer}(1998)}]{Hubert-book}
\bibinfo{author}{\bibfnamefont{A.}~\bibnamefont{Hubert}} \bibnamefont{and}
  \bibinfo{author}{\bibfnamefont{R.}~\bibnamefont{Sch\"{a}fer}},
  \emph{\bibinfo{title}{Magnetic Domains- The analysis of magnetic
  microstructures}} (\bibinfo{publisher}{Springer}, \bibinfo{year}{1998}).

\bibitem[{\citenamefont{Betz}(1997)}]{these-betz}
\bibinfo{author}{\bibfnamefont{J.}~\bibnamefont{Betz}}, Ph.D. thesis,
  \bibinfo{school}{Universit\'e Grenoble I} (\bibinfo{year}{1997}).

\bibitem[{\citenamefont{{\relax Le Gall}
  et~al.}(2001{\natexlab{a}})\citenamefont{{\relax Le Gall}, {\relax Ben
  Youssef}, Tiercelin, Preobrazhsensky, Pernod, and
  Ostorero}}]{LeGall_IEEE_2001}
\bibinfo{author}{\bibfnamefont{H.}~\bibnamefont{{\relax Le Gall}}},
  \bibinfo{author}{\bibfnamefont{J.}~\bibnamefont{{\relax Ben Youssef}}},
  \bibinfo{author}{\bibfnamefont{N.}~\bibnamefont{Tiercelin}},
  \bibinfo{author}{\bibfnamefont{V.}~\bibnamefont{Preobrazhsensky}},
  \bibinfo{author}{\bibfnamefont{P.}~\bibnamefont{Pernod}}, \bibnamefont{and}
  \bibinfo{author}{\bibfnamefont{J.}~\bibnamefont{Ostorero}},
  \bibinfo{journal}{IEEE Trans. Magn.} \textbf{\bibinfo{volume}{37}},
  \bibinfo{pages}{2699} (\bibinfo{year}{2001}{\natexlab{a}}).

\bibitem[{\citenamefont{Stoner and E.P.Wohlfarth}(1991)}]{SW}
\bibinfo{author}{\bibfnamefont{E.}~\bibnamefont{Stoner}} \bibnamefont{and}
  \bibinfo{author}{\bibnamefont{E.P.Wohlfarth}}, \bibinfo{journal}{IEEE Trans.
  Magn.} \textbf{\bibinfo{volume}{27}}, \bibinfo{pages}{3475}
  (\bibinfo{year}{1991}), \bibinfo{note}{reprint of Phil. Trans. Roy. Soc.
  London vol A 240 pp 599-642 (1948)}.

\bibitem[{\citenamefont{Thiaville}(1998)}]{Thiaville-SW}
\bibinfo{author}{\bibfnamefont{A.}~\bibnamefont{Thiaville}},
  \bibinfo{journal}{J. Magn. Magn. Mater.} \textbf{\bibinfo{volume}{182}},
  \bibinfo{pages}{5} (\bibinfo{year}{1998}).

\bibitem[{\citenamefont{{\relax Le Gall}
  et~al.}(2001{\natexlab{b}})\citenamefont{{\relax Le Gall}, {\relax Ben
  Youssef}, Tiercelin, Preobrazhsensky, and Pernod}}]{Legall-Jap}
\bibinfo{author}{\bibfnamefont{H.}~\bibnamefont{{\relax Le Gall}}},
  \bibinfo{author}{\bibfnamefont{J.}~\bibnamefont{{\relax Ben Youssef}}},
  \bibinfo{author}{\bibfnamefont{N.}~\bibnamefont{Tiercelin}},
  \bibinfo{author}{\bibfnamefont{V.}~\bibnamefont{Preobrazhsensky}},
  \bibnamefont{and} \bibinfo{author}{\bibfnamefont{P.}~\bibnamefont{Pernod}},
  \bibinfo{journal}{J. Magn. Soc. Japan} \textbf{\bibinfo{volume}{25}},
  \bibinfo{pages}{258} (\bibinfo{year}{2001}{\natexlab{b}}).

\bibitem[{\citenamefont{Prutton}(1964)}]{Prutton}
\bibinfo{author}{\bibfnamefont{M.}~\bibnamefont{Prutton}},
  \emph{\bibinfo{title}{Thin ferromagnetic films}}
  (\bibinfo{publisher}{Butterworths}, \bibinfo{year}{1964}).

\bibitem[{\citenamefont{Labrune et~al.}(1997)\citenamefont{Labrune, Kools, and
  Thiaville}}]{Labrune}
\bibinfo{author}{\bibfnamefont{M.}~\bibnamefont{Labrune}},
  \bibinfo{author}{\bibfnamefont{J.~C.~S.} \bibnamefont{Kools}},
  \bibnamefont{and}
  \bibinfo{author}{\bibfnamefont{A.}~\bibnamefont{Thiaville}},
  \bibinfo{journal}{J. Magn. Magn. Mater.} \textbf{\bibinfo{volume}{171}},
  \bibinfo{pages}{1} (\bibinfo{year}{1997}).

\bibitem[{\citenamefont{Chopra et~al.}(1999)\citenamefont{Chopra, Ludwig,
  Quandt, Hua, Brown, Swartzendruber, and Wuttig}}]{Chop-99}
\bibinfo{author}{\bibfnamefont{H.}~\bibnamefont{Chopra}},
  \bibinfo{author}{\bibfnamefont{A.}~\bibnamefont{Ludwig}},
  \bibinfo{author}{\bibfnamefont{E.}~\bibnamefont{Quandt}},
  \bibinfo{author}{\bibfnamefont{S.~Z.} \bibnamefont{Hua}},
  \bibinfo{author}{\bibfnamefont{H.~J.} \bibnamefont{Brown}},
  \bibinfo{author}{\bibfnamefont{L.}~\bibnamefont{Swartzendruber}},
  \bibnamefont{and} \bibinfo{author}{\bibfnamefont{M.}~\bibnamefont{Wuttig}},
  \bibinfo{journal}{J. Appl. Phys.} \textbf{\bibinfo{volume}{85}},
  \bibinfo{pages}{6238} (\bibinfo{year}{1999}).

\bibitem[{\citenamefont{Shull et~al.}(2004)\citenamefont{Shull, Quandt,
  Shapiro, Glasmachers, and Wuttig}}]{MS-MOIF}
\bibinfo{author}{\bibfnamefont{R.~D.} \bibnamefont{Shull}},
  \bibinfo{author}{\bibfnamefont{E.}~\bibnamefont{Quandt}},
  \bibinfo{author}{\bibfnamefont{A.}~\bibnamefont{Shapiro}},
  \bibinfo{author}{\bibfnamefont{S.}~\bibnamefont{Glasmachers}},
  \bibnamefont{and} \bibinfo{author}{\bibfnamefont{M.}~\bibnamefont{Wuttig}},
  \bibinfo{journal}{J. Appl. Phys.} \textbf{\bibinfo{volume}{95}},
  \bibinfo{pages}{6948} (\bibinfo{year}{2004}).

\end{thebibliography}

\end{document}